\documentclass[a4paper,twocolumn,11pt,accepted=2023-06-12]{quantumarticle}

\pdfoutput=1
\usepackage[utf8]{inputenc}
\usepackage[english]{babel}
\usepackage[T1]{fontenc}
\usepackage{amsmath}
\usepackage{hyperref}

\usepackage{tikz}
\usepackage{lipsum}

\usepackage{graphicx}
\usepackage{dcolumn}
\usepackage{bm}

\usepackage{amssymb}
\usepackage{bbm}
\usepackage{nicefrac}
\usepackage{braket}
\usepackage{latexsym}
\usepackage{soul}
\usepackage{braket}
\usepackage{cancel}
\usepackage{physics}

\begin{document}

\title{Qubit Number Optimization for Restriction Terms of QUBO Hamiltonians}

\author{Iñigo Perez Delgado}
\email{i.perez.delgado@ibermatica.com}
\author{Beatriz García Markaida}
\affiliation{i3b Ibermatica, Parque Tecnológico de Bizkaia, Ibaizabal Bidea, Edif. 501-A, 48160 Derio, Spain}
\author{Alejandro Mata Ali}
\affiliation{i3B Ibermatica, Quantum Development Department, Paseo Mikeletegi 5,  20009 Donostia, Spain}
\author{Aitor Moreno Fdez. de Leceta}
\affiliation{i3b Ibermatica, Unidad de Inteligencia Artificial, Avenida de los Huetos,
Edificio Azucarera, 01010 Vitoria, Spain}


\begin{abstract}
In usual restriction terms of the Quadratic Unconstrained Binary Optimization (QUBO) hamiltonians, a integer number of logical qubits R, called the Integer Restriction Coefficient (IRC), are forced to stay active. In this paper we gather the well-known methods of implementing these restrictions, as well as some novel methods that show to be more efficient in some frequently implemented cases. Moreover, it is mathematically allowed to ask for fractional values of $R$. For these Fractionary Restriction Coefficients (FRC) we show how they can reduce the number of qubits needed to implement the restriction hamiltonian even further. Lastly, we characterize the response of DWave's Advantage$\_$system4.1 Quantum Annealer (QA) when faced with the implementation of FRCs, and offer a summary guide of the presented methods and the situations each of them is to be used.
\end{abstract}

\maketitle

\section{\label{sec:intro}Introduction}
The idea of a quantum advantage, understood as a computational speedup achievable only by a quantum computer, was conceived independently in Yuri Manin’s 1980 book “Computable and Noncomputable” \cite{manin1980computable} and by Richard Feynman in his 1982 keynote speech “Simulating physics with Computers” \cite{feynman1982computer}. However, both these authors thought of quantum computers mainly as machines to simulate other quantum systems. Not much later, in 1985, the rigorous description of quantum Turing machines was published by David Deutsch \cite{Deutsch1985qcDescription}. This was the first instance of a quantum computer conceptualised not only as a quantum system simulator but as a general problem solving machine. Deutsch did not, though, show any computational advantage of quantum computers over classical ones. The first true algorithms to do so, Peter Shor’s factoring and discrete $\log$ algorithms, were published in 1994 \cite{Shor1994}, and reignited the field of Quantum Computing (QC).

The first experimental demonstration of a quantum algorithm came in 1998 by
Jones and Mosca \cite{Jones1998}. By the year 2000, the field was mature enough for a standard textbook such as “Quantum Computation and Quantum Information” \cite{Nielsen}, by Nielsen and Chuang, to appear. In 2001, the first quantum execution of Shor’s factoring algorithm was performed by Vandersypen et al. \cite{Vandersypen2001}.

Today, two main branches of QC exist. On the one hand we have those who follow Deutsch's path of logical gates and Turing-completeness, called gate-based QC. On the other hand there are quantum annealers, like 
the DWave’s Advantage system4.1 used in the simulations for this paper, which are more similar to Manin and Feynman's idea of an analog simulator or a digital twin. These annealers solve optimization problems by starting the machine from the ground state of some initial hamiltonian, which is slowly morphed into another hamiltonian. This second, target hamiltonian is constructed in such a way that its relaxed state corresponds with the answer of the maximization problem. Then, if the change has allowed the adiabatic theorem \cite{Born1928} to apply, the system will be in the desired relaxed state, and measuring it just once will give the answer. However, in today's Noisy Intermediate-Scale Quantum (NISQ) era quantum annealers the probability $P(E_0)$ of having the ground state is not always $100\%$, so for any interesting-sized problem multiple runs are needed.

Gate-based QC, as a Turing-complete method, can also solve these optimization problems via methods such as the Variational Quantum Eigensolver (VQE) and the Quantum Approximate Unconstrained Algorithm (QAOA). However, gate-based devices suffer the same imperfections as quantum annealers, as they also reach $P(E_0)<1$.

In this paper we will explore the different values of the restriction coefficients of the restriction hamiltonian, a crucial part of QUBO hamiltonians explained in Sec. \ref{sec:QUBO}. 
Then, in Sec. \ref{sec:integer}, we will explore the usage of Integer Restriction Coefficients (IRC) both for problems with a single minimum (SSec. \ref{ssec:integer1}) and with multiple minima (SSec. \ref{ssec:integerM} and SSec. \ref{ssec:integerMequi}). In Sec. \ref{sec:halfinteger} we caracterize the benefits of one simple half-integer method when applied to DWave’s
Advantage system4.1 device (SSec. \ref{ssec:Comparison}) and show the improvement of Fractionary Restriction Coefficients (FRC) over IRC for the general case of $M$ different values of $R$ (SSec. \ref{ssec:improveHI}). In Sec. \ref{sec:beyondhalfinteger} we show the response of the Advantage system4.1 device when faced with $R\in\mathbb{Z}$ beyond just half-integers. Lastly in Sec. \ref{sec:conclusions}, we collect the best presented methods for each case and give a summary of the findings of this work.


\section{QUBO hamiltonians\label{sec:QUBO}}

In order to facilitate the work of the hardware and achieve $P(E_0)$ as close to one as possible, target hamiltonians (which have the role of cost functions) are usually written in Quadratic Unconstrained Binary Optimization (QUBO) form. Being quadratic, they can contain at most terms of second order in the binary variables $x_i\in\{0,1\}$, that is, $C_{ij}x_ix_j\quad:C_{ij}\in\mathbbm{R}$. They can also contain lineal terms of the form $C_{i}x_i\quad:C_{i}\in\mathbbm{R}$ and a constant term $C\in\mathbbm{R}$. However, as $x_i\in\{0,1\}$, it happens that $x_i=x_i^2$, and so lineal terms can be rewritten as quadratic terms. The constant term does not affect the optimization problem and should be ignored when solving it. Then, effectively, all relevant terms of a QUBO hamiltonian are of the form $C_{ij}x_ix_j$, which allows the encoding of the whole $i\in\{1,N\}$ problem in a single $N\times N$ matrix.  

The terms of these hamiltonians are, for comprehensiveness, divided into two main blocks. The first of them, the `problem hamiltonian', encodes the equivalent of the cost function of the classical problem. The second one, the `restriction hamiltonian', encodes the equivalent of classical search-space limits as highly-penalised regions of the quantum search space. It is important to notice, however, that these restrictions are never absolute, since our problem is to be encoded into a QUBO hamiltonian, unconstrained by definition. The most usual terms of a restriction hamiltonians are IRC terms.

\section{\label{sec:integer}Integer Restriction Coefficients}

Usual restriction terms of the optimization hamiltonian have integer values for their restriction coefficients $R$. These IRCs are, of course, perfectly useful, and there are some cases in which FRCs do not suppose any improvement. Moreover, most things FRCs can do are also achievable by IRC methods, which makes them useful for comparison purposes.

\subsection{\label{ssec:integer1} Case with one single allowed value of $R$}

Let it be $\tilde{\Omega}=\{\tilde{x}_1, \tilde{x}_2,...\}$ the set of all
 binary variables of a minimization optimization problem, where each $\tilde{x}_i$ binary variable can only take the values of $0$ or $1$. In order to have, for a subset of $\tilde{\Omega}$ of size $N$ denoted as $\Omega=\{x_1, x_2,...,x_N\}$, that for some integer value of $R\geq 0$
\begin{equation}
    \quad \sum_{\Omega} x_{i}=R \;,
    \label{eq:integer_condition}
\end{equation}
 we introduce a term $H$ in the cost function of the form

\begin{equation}
    H = \lambda\left(\sum_{\Omega} x_{i}- R\right)^2\;.
    \label{eq:R_at_most}
\end{equation}

That term $H\geq 0$ will only equal zero if Eq.  \ref{eq:integer_condition} is fulfilled. For large enough values of the Lagrange multiplier $\lambda>0$, combinations of $x_{i}$ that do not follow that condition will be prohibited from being the ground state.

\subsection{\label{ssec:integerM}General case of $M$ possible values of $R$}

If multiple values of the restriction coefficient $R$ are to be allowed, a number of $y_k$ dummy variables can be introduced. For an ordered set $\{R_1, R_2,..\}$ where $R_k<R_l \Leftrightarrow k<l$, if

\begin{equation}
    \quad \sum_{\Omega} x_{i}=R\in \{R_1, R_2,...,R_{M}\}
    \label{eq:integer_condition_multiple_Rs}
\end{equation}
is to be fulfilled, then Eq. \ref{eq:R_at_most} can be generalized to

\begin{equation}
    H = \lambda_1\left(\sum_{\Omega} x_{i}-\sum_{k=1}^{M} R_k\,y_k\right)^2
    +\lambda_2\left(\sum_{k=1}^{M} y_k- 1\right)^2\;,
    \label{eq:generalM}
\end{equation}
for two large positive Lagrange multipliers $\lambda_1$ and $\lambda_2$. Notice how the second term has the shape of Eq. \ref{eq:R_at_most}, forcing the activation of only one $y_k$. Using only integer restriction coefficients, for $M$ possible values of $R$, $M$ dummy variables are needed. 


\subsection{\label{ssec:integerMequi}Case with $M$ equispated values of $R$}

In the special case of equispatially valued $R$, where $\forall k\; R_k=R_1+\Delta(k-1)$ for some gap between values $ \Delta\geq 0$, another strategy can be followed, using only $M-1$ dummy variables. This is shown in Eq. \ref{eq:linealM}.

\begin{equation}
    H = \lambda\left(\sum_{\Omega} x_{i}+\Delta\sum_{k=1}^{M-1} \,y_k- R_{M}\right)^2\;.
    \label{eq:linealM}
\end{equation}

However, there is a more efficient way of writing that term, which uses only $\mathcal{O}(\log_2M)$ dummy variables, as shown in Eq. \ref{eq:logM}.

\begin{equation}
    H = \lambda\Bigg(\sum_{\Omega} x_{i}-\Delta\sum_{n=0}^{\gamma-1} \,2^n y_n - \Delta\left(M-2^{\gamma}\right)y_\gamma -R_1\Bigg)^2\;, 
    \label{eq:logM}
\end{equation}
where
\begin{equation}
    \gamma = \lfloor\log_2M\rfloor\;. 
\end{equation}

As can be seen, only $N_y=\gamma+1$ extra variables are used in general. When $M$ is a power of 2, the coefficient of $y_\gamma$ equals zero, which reduces the number of required dummy variables to just $N_y=\gamma$.

\section{\label{sec:halfinteger}Half-integer FRCs}

As $\forall i\; x_i\in\{0,1\}$, any sum of the form $\sum x_i$ must necessarily be an integer too. Therefore, one can never fullfill Eq. \ref{eq:integer_condition} for half-integer $R=n+0.5\;:n\in \mathbb{N}$. However, that will be the case for all states, whether they are wanted answers or not. This residual energy does not, as usual, affect the existence of minima.  In fact, as a half-integer $R<N$ is equidistant from the two allowed values of $\sum x_i=n$ and $\sum x_i=n+1$, those two consecutive values will appear as minima. 

\subsection{\label{ssec:ssec:improveHI3}Variable reduction method for consecutive $M=2$ case }

The degeneration of the energies of the minima of half-integer $R$ hamiltonian allows us to solve the $\Delta=1$, $M=2$ case of Eq. \ref{eq:logM} with no  dummy variables, instead of $\gamma=1$, as shown in the integer $R$ case of Eq. \ref{eq:M=2halfinteger}.

\begin{equation}
    H = \lambda\left(\sum_{\Omega} x_{i}-n-0.5\right)^2\;.
    \label{eq:M=2halfinteger}
\end{equation}

The minima of the hamiltonian of Eq. \ref{eq:M=2halfinteger} are those integer values of $\sum x_i$ which are closest from $n+0.5$, that is, $n$ and $n+1$, as intended.

One of the many uses of the $n=0$ version of this method, which allows for $\sum x_i\in\{0,1\}$, is the resolution of the Job Selection Problem, a generalization of the well-known Travelling Salesperson Problem. The DWave implementation of that practical case can be found in \cite{Hobbit2022}.

\subsection{\label{ssec:Comparison}Comparison between integer and half-integer methods}

We have compared, using the Advantage$\_$system4.1 DWave quantum annealer, the performance of the integer method $R\in\{n,n+1\}$ method of Eq. \ref{eq:logM} and the half-integer $R=n+0.5$ method of Eq. \ref{eq:M=2halfinteger} for the $n=1$ cases of $N\in\{3,15\}$. The two resulting hamiltonians are shown in Eq. \ref{eq:logMcompare} and Eq. \ref{eq:M=2halfintegercompare}.

\begin{equation}
   R\in\{1,2\}\quad\Rightarrow\quad H = \lambda\Bigg(\sum_{i=1}^{N} x_{i}- y_0-1\Bigg)^2\;,
    \label{eq:logMcompare}
\end{equation}

\begin{equation}
  R=1.5 \quad\Rightarrow\quad   H = \lambda\left(\sum_{i=1}^N x_{i}-1.5\right)^2\;.
    \label{eq:M=2halfintegercompare}
\end{equation}

.

\begin{figure}[htb]
  \includegraphics[width=1\linewidth]{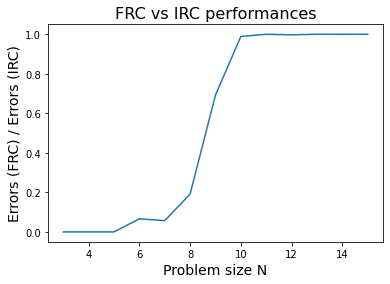}
  \caption{Error comparison for hamiltonians of Eq. \ref{eq:logMcompare} and Eq. \ref{eq:M=2halfintegercompare}. For all sizes of the problem, the FRC method behaves better or equal than the IRC method.}
  \label{fig:comparison}
\end{figure}

As seen in Fig. \ref{fig:comparison}, for all measured values, the hamiltonian with half-integer $R$ of Eq. \ref{eq:M=2halfintegercompare} had a smaller or equal number of answers with $\sum x_i\notin\{1,2\}$. This means that not only the half-integer method uses less resources, but it results in more accurate answers, specially when dealing with subsets $\Omega$ of size 9 or lower.

\subsection{\label{ssec:improveHI3}Variable reduction method for consecutive $M$=3 case}

For the $M=3$ case of $\sum x_i\in\{n,n+1,n+2\}$, one can use

\begin{equation}
    H = \lambda\left(\sum_{\Omega} x_{i}-y_1-n-0.5\right)^2\;.
    \label{eq:M=3halfinteger}
\end{equation}

This can be generalized to

\begin{equation}
    H = \lambda\left(\sum_{\Omega} x_i-\sum_{k=1}^{M-2} y_k-n-0.5\right)^2\;,
    \label{eq:Mhalfinteger}
\end{equation}

but $N_y=M-2$ dummy variables are needed, vs. the $\mathcal{O}(\log_2M)$ needed using Eq. \ref{eq:logM}. This makes the generalization not useful for any $M>3$, as shown in Tab. \ref{tab:escalation}.

\begin{table}[htb]
    \centering
    \begin{tabular}{|c|c|c|c|c|c|c|}
    \hline
       $M$& $2$& $3$& $4$& $5$& $6$& $7$\\
    \hline
       $N_y$ from Eq. \ref{eq:Mhalfinteger} & $0$& $1$& $2$& $3$& $4$& $5$\\
    \hline
       $N_y$ from Eq. \ref{eq:logM} & $1$& $2$& $2$& $3$& $3$& $3$\\
    \hline  
    \end{tabular}
    \caption{Escalation of the two methods as a function of $M$. The $\mathcal{O}(M)$ method is the most efficient only for $M=2$ and $M=3$. For $M=4$ and $M=5$ it requires the same $N_y$ number of dummy variables as the $\mathcal{O}(\log_2M)$ method, and from $M=6$ onwards it becomes less efficient.}
    \label{tab:escalation}
\end{table}

\subsection{\label{ssec:improveHI}Variable reduction method for the general $M$-valued case}

For the general case of Eq. \ref{eq:generalM}, where the possible values of $\sum x_i$ are not equispaced, using half-integer values of $R$ reduces the number of needed dummy variables from  $M$ to only $M-1$, as seen in Eq. \ref{eq:generalMreduced}.

\begin{equation}
\begin{split}
    H = \lambda_1&\left(\sum_{\Omega} x_{i}-\sum_{k=2}^{M} (R_k-R_1)\,y_k-R_1\right)^2\\ &+\lambda_2\left(\sum_{k=2}^{M} y_k- 0.5\right)^2\;.
\end{split}
    \label{eq:generalMreduced}
\end{equation}

The second term of Eq. \ref{eq:generalMreduced} allows for one or none of the $y_k$ variables to be on. If none are, then the first term makes $\sum x_i=R_1$. On the other hand, if only the $k$-th variable is $y_k=1$, then the first term makes $\sum x_i=(R_k-R_1)+R_1 \Rightarrow \sum x_i=R_k$. As only $k\in\{2,M\}$ are used, only $M-1$ variables are needed.

\section{\label{sec:beyondhalfinteger}Beyond half-integer FRCs}
As we have seen in Sec. \ref{sec:halfinteger}, not only integer values of $R$ have a use. After analizing the response of the Quantum Annealer for half-integer values, the next natural step is to try general fractional values beyond half-integers. 

Ideally, an annealer would have as its only output the ground state of the system: the probability distribution of the measurement would be a Kronecker delta $P(E)=\delta_{E,E_0}/M$, where $E_0$ is the energy of each of the $M$ ground states and $E$ the energy of the evaluated system. Then, as the only fractionary values which are equidistant from two integers are, by definition, half-integers, having a general fractionary value for $R$ should have the same effect as using the closest integer value. For example, if $R=6.4$, then following Eq. \ref{eq:R_at_most} $\sum x_i=6$ would be the ground state with an energy of $E=E_0=0.16\lambda$, followed by $\sum x_i=7$ with $E=0.36\lambda$, $\sum x_i=5$ with $E=1.96\lambda$ and so on. 

However, when the noise of physical annealers is taken into account, the probability of measuring any state becomes some general function $P(E)=f(\vert E-E_0\vert )$ where in principle $\frac{\partial f}{\partial \vert E-E_0\vert}\leq0$. The exact profile of $f$ will depend on the physical inner workings of the device itself.

\begin{figure}[htb]
  \includegraphics[width=1\linewidth]{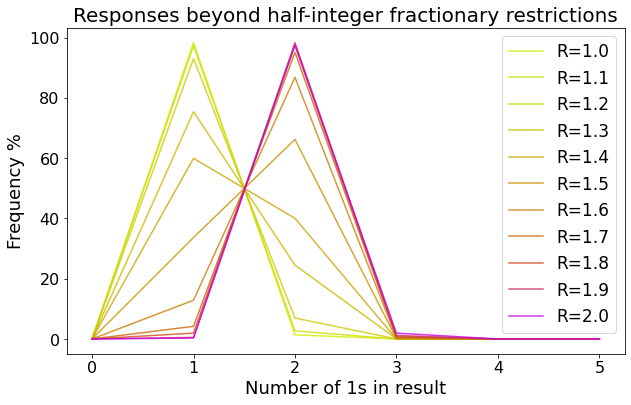}
  \caption{Evolution of $f_R$ for $R\in[1,2]$ in a 5-qubit system. As expected, for $R=1$ almost $100\%$ of the measurements give $\sum x_i=1$, and for $R=2$ almost $100\%$ of them give $\sum x_i=2$. For increasing intermediate values of $R$, the measurement probability migrates from $\sum x_i=1$ to $\sum x_i=2$. The percentage of results with $0$, $3$, $4$ and $5$ active qubits is close to $0\%$ for all presented values of $R$.}
  \label{fig:multiR}
\end{figure}

For the Advantage$\_$system4.1 DWave quantum annealer, we conducted eleven 5-qubit demonstrations with $R\in[1,2]$ to obtain the probabilities of measuring the six allowed values for $\sum x_i\in\{0,1,2,3,4,5\}$. The obtained profile $f_R$ varies continuously, always having its maxima in the integer which mimimizes $\sum x_i-R$, as can be seen in Fig. \ref{fig:multiR}. 
 
\begin{figure}[htb]
  \includegraphics[width=1\linewidth]{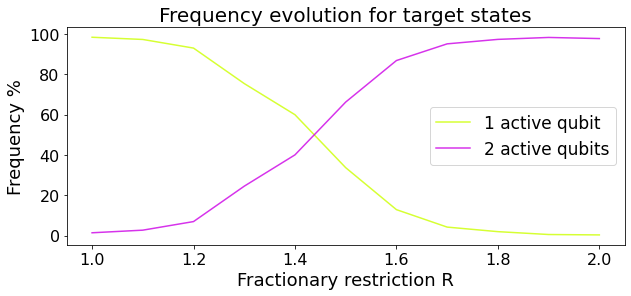}
  \caption{Probability transference between $\sum x_i=1$ and $\sum x_i=2$ as $R$ changes from $1$ to $2$. In an ideal device, the transference should look like a step function centered in $M=1.5$. However, the measured profile is closer to a cosine than to a steep transference.}
  \label{fig:swap}
\end{figure}

In order to better appreciate the probability transfer between $\sum x_i=1$ and $\sum x_i=2$, since $\forall R$ we measure $P(\sum x_i)\approx0$ for all other values of $\sum x_i$, we have plotted Fig. \ref{fig:swap}.

\section{\label{sec:conclusions}Conclusions}

In this paper we have identified the advantages of half-integer Fractionary Restriction Coefficients of QUBO restriction hamiltonians. These FRC methods, along with several Integer Restriction Coefficient methods described in the paper, are gathered to offer a comprehensible guide of dummy variable number reduction methods for QUBO restriction hamiltonians. These methods are, as usual, translatable to Ising hamiltonians as well. 

For each restriction form, the most useful method has been gathered in Tab. \ref{tab:summary}, stating its IRC/FRC nature and the needed $y_k$ number of dummy variables as a function of the $M$ number of discrete possible values the restriction accepts as minima of the cost function. For the most simple and well-known method $y_k=M-1$, while some of the described methods achieve the exponential advantage of dummy variable numbers of only $y_k=\mathcal{O}(\log_2M)$.

\begin{table}[]
    \centering
    \begin{tabular}{|c|c|c|}
    \hline
        CASE & Eq. & NEEDED $y_k$\\
    \hline
        \begin{tabular}{@{}c@{}}$M=1$, \\ single allowed value\end{tabular} & \ref{eq:R_at_most} &$0$\\
    \hline
        \begin{tabular}{@{}c@{}}$M=2$, \\$\sum{x_i} \in \{n, n+1\}$\end{tabular}  &\ref{eq:M=2halfinteger} &  $0$\\
    \hline
        \begin{tabular}{@{}c@{}}$M=3$, \\$\sum{x_i} \in \{n, n+1,n+2\}$\end{tabular}   & \ref{eq:M=3halfinteger} &  $1$\\
     \hline
        \begin{tabular}{@{}c@{}}$R_{k+1}-R_k=\Delta$ \\$:\log_2M\notin \mathbbm{N}$\end{tabular} &\ref{eq:logM} &  $\lfloor\log_2M\rfloor+1$\\
    \hline
        \begin{tabular}{@{}c@{}}$R_{k+1}-R_k=\Delta$ \\$:\log_2M\in \mathbbm{N}$\end{tabular}  & \ref{eq:logM} &  $\lfloor\log_2M\rfloor$\\
    \hline
        \begin{tabular}{@{}c@{}}General case \\ $R\in \{R_1, R_2,...,R_{M}\}$\end{tabular}  & \ref{eq:generalMreduced} &  $M-1$\\
      \hline  
    \end{tabular}
    \caption{Suggested optimal constructions of hamiltonian restriction terms. For low values of $M$, some of these expressions are equivalent. Those simple, often used cases have been included for completion.}
    \label{tab:summary}
\end{table}

\medskip

We have also characterised the response of the Advantage$\_$system4.1 DWave quantum annealer to general, non-half-integer FRC and found that the evolution of the measurement probability of states close to the value of the FRC $R$ has a non-neglegible difference with the step function that would be measured in the case of the ideal device. A function distance between the measured curve and that ideal step function could be used as a quality benchmark for both Quantum and Digital Annealers.

\section*{\label{sec:acknowledgments}Acknowledgments}
The research leading to this paper has received funding from the Q4Real project (Quantum Computing for Real Industries), HAZITEK 2022, no. ZE-2022/00033.

\section*{\label{sec:interests}Competing interests}
The authors declare no competing interests. We acknowledge use of the DWave for this work. The views expressed are those of the authors and do not reflect the official policy or position of DWave or the DWave team.

\bibliographystyle{quantum}
\bibliography{quantum}

\end{document}